\documentclass[aps,prl,twocolumn,letterpaper,superscriptaddress,floatfix,showpacs]{revtex4-1}
\usepackage{graphicx,amsmath,amsfonts,amssymb,color,ulem,bm,tabularx,sidecap,dcolumn,multirow,hyperref}
\usepackage{times}
\usepackage[sort&compress]{natbib}
\usepackage[T1]{fontenc}
\DeclareFontFamily{OT1}{pzc}{}
\DeclareFontShape{OT1}{pzc}{m}{it}{<-> s * [1.2] pzcmi7t}{}
\DeclareMathAlphabet{\mathpzc}{OT1}{pzc}{m}{it}
\newcommand{\ham}{\mathpzc{H}}
\newcommand{\BaCo}{Ba$_\text{3}$CoSb$_\text{2}$O$_\text{9}$}
\newcommand{\Co}{Co$^{2+}$}

\begin{document}

\title{Static and Dynamical Properties of the Spin-1/2 Equilateral Triangular-Lattice Antiferromagnet {\BaCo}}

\author{J. Ma}
\affiliation{Department of Physics and Astronomy, University of Tennessee, Knoxville, Tennessee 37996, USA}
\affiliation{Quantum Condensed Matter Division, Oak Ridge National Laboratory, Oak Ridge, Tennessee 37831, USA}

\author{Y. Kamiya}
\affiliation{iTHES Research Group and Condensed Matter Theory Laboratory, RIKEN, Wako, Saitama 351-0198, Japan}

\author{Tao Hong}
\author{H. B. Cao}
\author{G. Ehlers}
\author{W. Tian}
\affiliation{Quantum Condensed Matter Division, Oak Ridge National Laboratory, Oak Ridge, Tennessee 37831, USA}

\author{C. D. Batista}
\affiliation{Theoretical Division, T-4 and CNLS, Los Alamos National Laboratory, Los Alamos, New Mexico 87545, USA}

\author{Z. L. Dun}
\affiliation{Department of Physics and Astronomy, University of Tennessee, Knoxville, Tennessee 37996, USA}

\author{H. D. Zhou}
\affiliation{Department of Physics and Astronomy, University of Tennessee, Knoxville, Tennessee 37996, USA}
\affiliation{National High Magnetic Field Laboratory, Florida State University, Tallahassee, Florida 32310-3706, USA}

\author{M. Matsuda}
\affiliation{Quantum Condensed Matter Division, Oak Ridge National Laboratory, Oak Ridge, Tennessee 37831, USA}

\date{\today}
\begin{abstract}

  We present single-crystal neutron scattering measurements of the spin-1/2 equilateral triangular-lattice antiferromagnet {\BaCo}. Besides confirming that the {\Co} magnetic moments lie in the $ab$ plane for zero magnetic field and then determining all the exchange parameters of the minimal quasi-2D spin Hamiltonian, we provide conclusive experimental evidence of magnon decay through observation of intrinsic line-broadening. Through detailed comparisons with the linear and nonlinear spin-wave theories, we also point out that the large-$S$ approximation, which is conventionally employed to predict magnon decay in noncollinear magnets, is inadequate to explain our experimental observation. Thus, our results call for a new theoretical framework for describing excitation spectra in low-dimensional frustrated magnets under strong quantum effects.

\end{abstract}

\pacs{61.05.F-, 75.10.Jm, 75.45.+j, 78.70.Nx}

\maketitle

\textit{Introduction.---}%
The $S=1/2$ triangular-lattice Heisenberg antiferromagnet (TLHAF) is the paradigmatic example of a two-dimensional (2D) frustrated quantum magnet~\cite{Balents2010,Moessner2006,Anderson1973,Fazekas1974,Mezio2012,Zheng2006,Starykh2006,Chernyshev2006,Chernyshev2009,Mourigal2013,Yamamoto2014,Yamamoto2015,Sellmann2015,Ghioldi2015}. The combination of frustration, strong quantum fluctuations and low dimensionality is anticipated to produce strong deviations from semiclassical theories. While several distorted triangular-lattice materials, such as $\kappa$-(BEDT- TTF)$_{2}$Cu$_{2}$(CN)$_{3}$~\cite{Shimizu2003}, Cs$_{2}$Cu$X_{4}$ ($X$ = Cl~\cite{Coldea2003,Breunig2015,vanWell2015} and Br~\cite{Ono2003,Fortune2009,vanWell2015}), and CuCrO$_{2}$~\cite{Poienar2010,Frontzek2011}, have been investigated in the past, the distorted structures introduce additional terms, such as the Dzyaloshinskii-Moriya (DM) interaction, into the paradigmatic Hamiltonian ~\cite{Coldea2003,Ono2003,Fortune2009,Starykh2010,Breunig2015,vanWell2015}.

The \textit{equilateral} triangular-lattice quantum antiferromagnet {\BaCo} was synthesized recently~\cite{Doi2004,Shirata2012,Zhou2012,Susuki2013,Koutroulakis2015,Quirion2015}. The {\Co} ion has a Kramers doublet ground-state due to the spin-orbit coupling, and this doublet can be described as an effective spin-1/2 moment. In addition, the high symmetry of the hexagonal crystal structure, {\it P}6$_{3}$\,/\,{\it mmc}~\cite{Doi2004, Shirata2012, Zhou2012, Susuki2013, Koutroulakis2015}, forbids DM interaction for pairs up to third nearest-neighbor (NN) in the same $ab$-plane and between any pair of spins along the $c$-axis~\cite{Shirata2012}. 

Powder neutron diffraction measurements presented the noncollinear 120$^\circ$ structure with the magnetic wavevector {\bf Q} = (1/3,\,1/3,\,1)~\cite{Doi2004}. The N\'{e}el temperature was found to be $\approx$\,3.8\,K and a rich temperature-magnetic field phase diagram was reported up to 32\,T~\cite{Shirata2012,Zhou2012,Susuki2013,Koutroulakis2015}. Electronic spin resonance (ESR)~\cite{Susuki2013} and nuclear magnetic resonance (NMR)~\cite{Koutroulakis2015} measurements suggested a spin model with small easy-plane  exchange anisotropy and an exchange interaction along the $c$-axis much weaker than the NN intralayer exchange. This observation is consistent with the alternation of  magnetic ({\Co})   and  nonmagnetic (Sb$_{2}$O$_{9}$ bioctahedra) layers along the $c$-direction. While more precise determination of the model parameters requires inelastic neutron scattering (INS) measurements, such detailed information is indeed physically relevant. The reason is that, according to the semiclassical theories in Refs.~\onlinecite{Chernyshev2006,Chernyshev2009}, ``smoking gun'' features of the magnetic excitations of the 2D $S$=1/2 TLHAF, such as the line-broadening of the single-magnon excitations resulting from spontaneous magnon decays~\cite{Starykh2006,Chernyshev2006,Chernyshev2009,Zhitomirsky2013,Mourigal2013}, can be rather sensitive to small deviations from the ideal model.

In this Letter, we present direct evidence of the magnetic structure in {\BaCo} and the detailed profile of magnon excitations obtained by neutron scattering measurements. We confirm the 120$^\circ$ order lying in the $ab$ plane at zero-field~\cite{Susuki2013,Koutroulakis2015} and determine the exchange constants of the minimal quasi-2D XXZ Hamiltonian proposed in Refs.~\onlinecite{Susuki2013,Koutroulakis2015,Yamamoto2015}. The INS spectrum also exhibits intrinsic line-broadening. By comparing the INS profile against the linear spin-wave (LSW) theory and the LSW+$1/S$ corrections, we show that the quantum fluctuations produce a strong renormalization of the magnon dispersion as previous theoretical predictions~\cite{Zheng2006,Starykh2006}. More importantly, however, our thorough examination reveals that the semiclassical (i.e., large $S$) treatment is inadequate to explain the observed magnon decay and the associated line-broadening in this spin-1/2 system, thereby pointing to a need for developing an alternative theoretical framework. Thus, although quantum fluctuations are not enough to destroy magnetic ordering in {\BaCo},  our study indicates that these fluctuations are qualitatively modifying the excitation spectrum relative to the semi-classical (large -$S$) regime.

\begin{figure}
  \centering
  \includegraphics[width=0.9\hsize,bb=0 0 540 475]{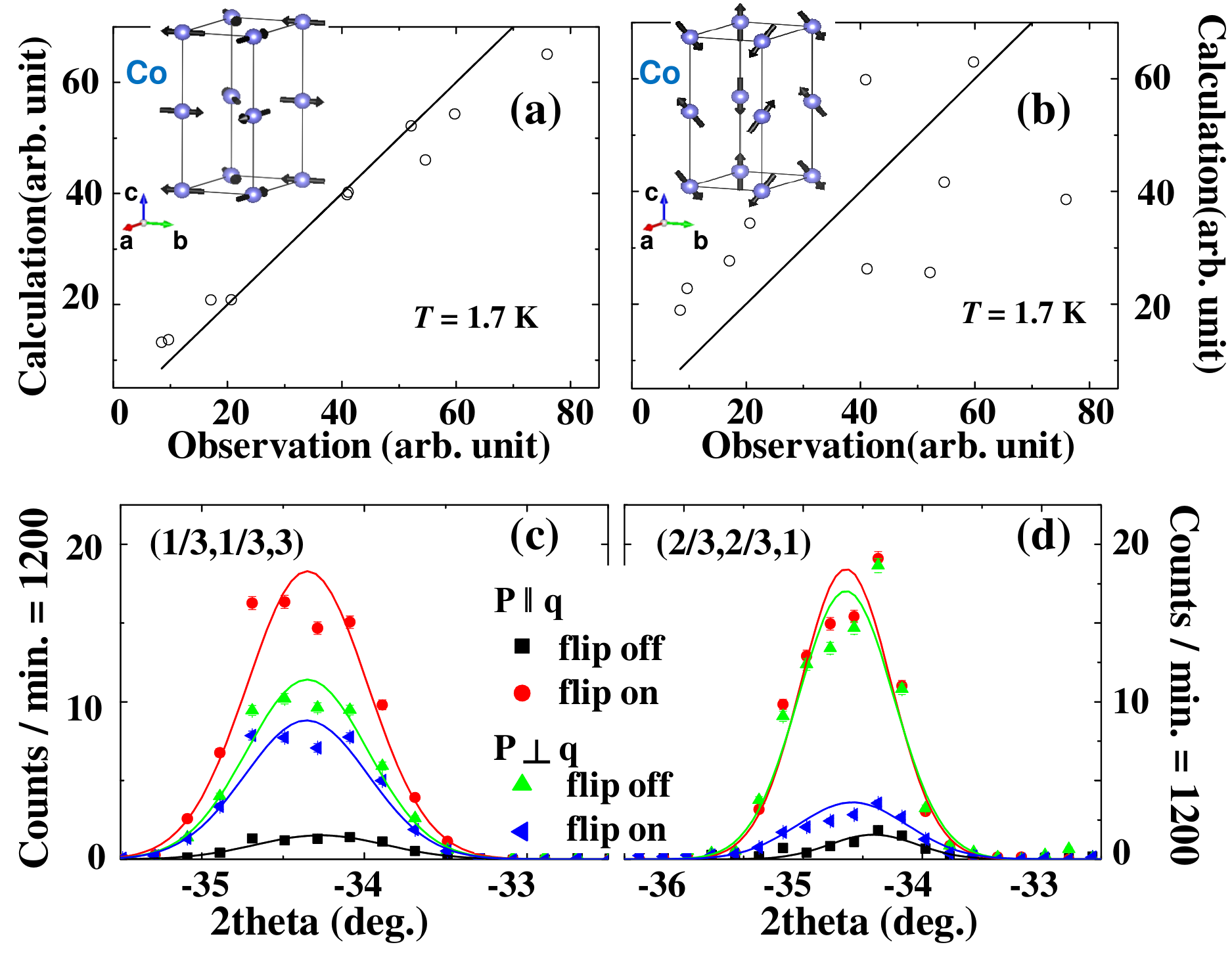}
  \caption{\label{fig:diff}%
    (color online)
    Comparison between the observed magnetic Bragg peak intensities at 1.7\,K on the HB-1A TAS and the simulated ones based on the 120$^\circ$ structure in the (a) $ab$ plane and (b) $ac$ plane. The solid lines are guides to the eye. Magnetic Bragg peaks at (c) (1/3,\,1/3,\,3) and (d) (2/3,\,2/3,\,1) measured with polarized neutrons at 1.7\,K on HB-1. The backgrounds have been subtracted. {\bf{P}} and {{\bf q}} are the polarization and the scattering vector, respectively.
  }
  \vspace{-10pt}
\end{figure}

\textit{Experiments.---}%
The single crystal of {\BaCo} [$\sim$1g, diameter (5mm)$\times$length (15mm)] was grown by the floating-zone technique and oriented in the (HHL) scattering plane for INS measurements. The unpolarized neutron diffraction data were obtained using the HB-1A triple-axis spectrometer (TAS) and the HB-3A four circle diffractometer at High Flux Isotope Reactor (HFIR), Oak Ridge National Laboratory (ORNL) \cite{supplemental}. The absence of site-disorder between Co and Sb was confirmed within an error of 1$\%$ and the magnetic wavevector ({\bf Q}) is (1/3,\,1/3,\,1) ~\cite{Doi2004}. We collected twenty magnetic Bragg peaks at 1.7\,K, and refined the data with FULLPROF~\cite{Rodriguez1993}. Two variants of the 120$^\circ$ structure were compared as the ``$ab$'' and ``$ac$'' plane models, which are favored by easy-plane and easy-axis anisotropy, respectively~\cite{Miyashita1986,Yamamoto2014,Yamamoto2015}. The $ab$  plane model corresponds to a 120$^\circ$ structure with all spins in the $ab$ plane [Fig.~\ref{fig:diff}(a)], while the $ac$ plane model assumes the 120$^\circ$ structure with one third of the moments parallel (or antiparallel) to the $c$-axis [Fig.~\ref{fig:diff}(b)]. Although the same magnetic Bragg peaks were generated from both structures, the related scattering intensity profiles were different. Our data are consistent with the $ab$ plane model, Figs.~\ref{fig:diff}(a) and (b).

We measured a series of  magnetic Bragg peaks, (1/3,\,1/3,\,1), (2/3,\,2/3,\,1), (1/3,\,1/3,\,3), and (1/3,\,1/3,\,5), and nuclear peaks, (1,\,1,\,0) and (0,\,0,\,6), using the HB-1 polarized neutron TAS at HFIR, ORNL for further confirmation of the spin directions. Figures~\ref{fig:diff}(c) and (d) show the (1/3,\,1/3,\,3) and (2/3,\,2/3,\,1) peaks, respectively. Since they are observed in the spin-flip channel with the initial polarization vector {\bf P} parallel to the scattering vector {\bf q}, both peaks are magnetic. More information on the magnetic order can be obtained by the neutrons polarized along the [1$\bar{1}$0] direction, which is perpendicular to the (HHL) plane, and then evaluating the intensity ratio $I_\text{sf}/I_\text{nsf}$ between the spin-flip and non-spin-flip channels. In our configuration, the spin-flip (non-spin-flip) scattering originates from the in-plane (out-of-plane) spin components. Through this analysis, the $ab$ plane model should be correct (Table~\ref{table:polar_HB1}). This confirms the easy-plane anisotropy proposed in the ESR~\cite{Susuki2013} and NMR~\cite{Koutroulakis2015} measurements in disagreement with Refs.~\onlinecite{Doi2004,Shirata2012,Zhou2012}~\cite{note}.

\begin{table}
  \caption{\label{table:polar_HB1}%
    Ratios between the spin-flip and non-spin-flip scattering intensities measured with neutrons polarized perpendicular to the scattering wavevector {\bf q}.
  }
  {
    \vspace{5pt}
    \renewcommand{\arraystretch}{1.2}
    \setlength{\tabcolsep}{6pt}
    \begin{tabular}{cccc}
      \hline
      \multirow{2}{*}{index} & \multirow{2}{*}{$I_\text{sf} /I_\text{nsf}$}&\multicolumn{2}{c}{magnetic model calculations}\\
      \cline{3-4}
      &  & $ab$ plane model & $ac$ plane model \\ \hline
      (2/3,2/3,1) &       0.16(2)  &   0.12 &  0.88  \\ \hline    
      (1/3,1/3,1) &       0.36(3)  &   0.33 &  0.67  \\ \hline     
      (1/3,1/3,3) &       0.81(2)  &   0.82 &  0.18  \\ \hline
      (1/3,1/3,5) &       0.94(2)  &   0.92 &  0.08  \\ \hline
    \end{tabular}
    \vspace{-5pt}
  }
\end{table}

\begin{figure*}
  \centering
  \includegraphics[width=0.9\hsize]{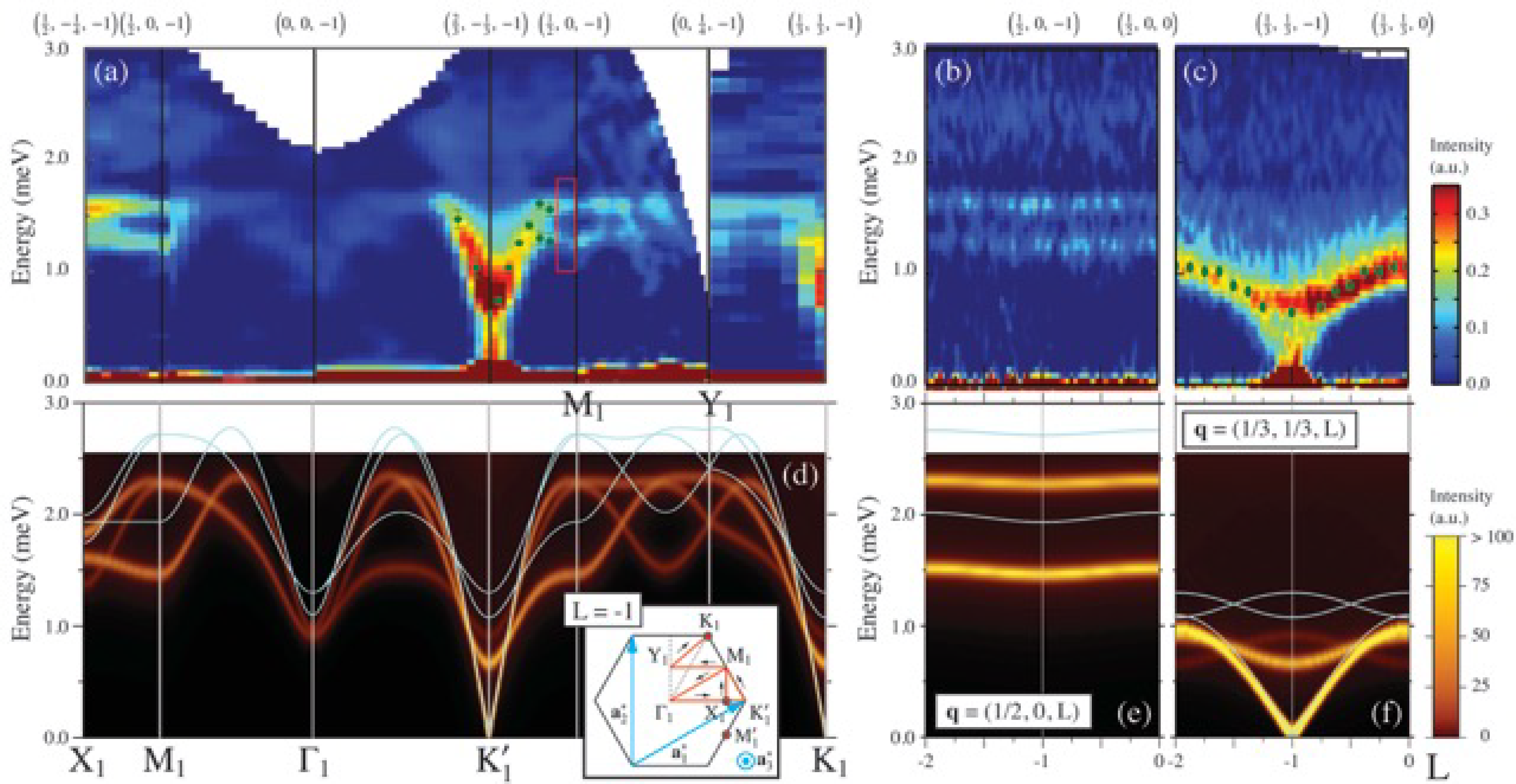}
  \caption{\label{fig:inel}%
    (color online)
    INS spectra of {\BaCo} as a function of the momentum and energy transfer at $T = 1.5\,\text{K}$ along the high symmetric (a) intralayer directions and the inter-layer (b) (1/2,\,0,\,L) and (c) (2/3,\,-1/3,\,L) directions in the reciprocal space. The background has been subtracted. The filled circles are peak positions from the measurements at the CG-4C TAS. The red rectangular frame in (a) represents the region where the decay effect is distinct and the details are discussed in Fig.3(a). (d)--(f) The intensity plot of the dynamical structure factor along the same symmetry lines as in (a)--(c) for $J = \text{1.7\,meV}$, $J'/J = \text{0.05}$, and $\Delta = \text{0.89}$ at $T = 0$ calculated by the nonlinear spin-wave approximation. The energy resolution (0.063 meV) has been convoluted. The solid lines represent the poles in the LSW approximation.
  }
  \vspace{-5pt}
\end{figure*}

To investigate the spin dynamics in {\BaCo}, we performed INS measurements on CG-4C cold neutron TAS at HFIR and Cold Neutron Chopper Spectrometer (CNCS) at Spallation Neutron Source (SNS), ORNL. At the CG-4C TAS, the final energy was fixed at either 5 or 3.5\,meV. The incident energy on CNCS was fixed at 3.315\,meV. Figures~\ref{fig:inel}(a)--\ref{fig:inel}(c) show the scattering profile at 1.5\,K along high-symmetry directions in the reciprocal space. Three dominant modes are observed as expected for the 120$^\circ$ structure. To be consistent with the theoretical calculation, Miller indices in following texts are labelled by the model notation.

\begin{figure}
  \centering
  \includegraphics[width=0.95\hsize,bb=0 0 158 256,trim=0 2 0 0,clip=true]{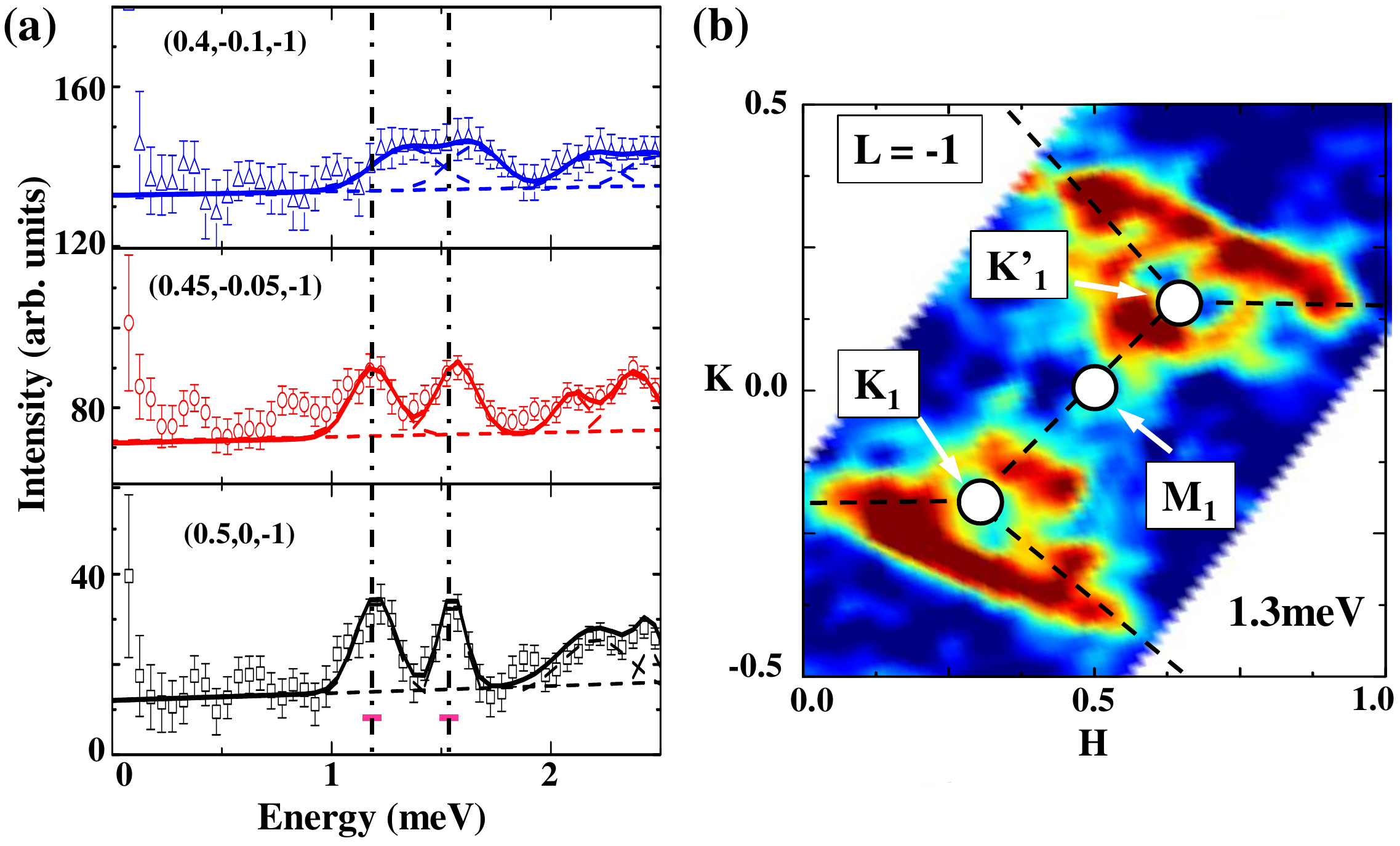}
  \caption{\label{fig:constQ-constE}%
    (color online)
      (a) The constant-{\bf q} scans near $\text{M}_1$ point, {\bf q} = (1/2,\, 0\,, -1). The horizontal error bars indicate the instrumental resolution. Thin dashed lines indicate individual fitted Gaussian peaks and background, and the solid line is their sum. The very broad feature above 2\,meV is attributed to the two-magnon continuum. (b) The constant-energy cut at 1.3\,meV as a function of {\bf q} = $(\text{H},\text{K},\,\text{L}=-1)$. The dashed lines are the BZ boundaries.
  }
  \vspace{-10pt}
\end{figure}

The overall bandwidth of discernible single-magnon branches is around 1.7\,meV for the in-plane dispersion. While our INS measurements resolved the finite bandwidth $\approx$\,1.1\,meV along the $c$-axis for the Goldstone mode [Fig.~\ref{fig:inel}(c)], which implies a  nonnegligible  inter-layer (antiferromagnetic) exchange. The relevance of the finite inter-layer exchange was pointed out recently~\cite{Koutroulakis2015,Yamamoto2015} to explain a weak anomaly in the magnetization curve at around 22\,T for a magnetic field $\mathbf{B}$ parallel to the $ab$-plane~\cite{Susuki2013}. The corresponding anomaly was clearly observed by NMR~\cite{Koutroulakis2015}. The gap of the quadratic band at {\bf q} = (2/3,\,-1/3,\,-1) is around 0.65\,meV, which agrees well with 170\,GHz by ESR~\cite{Susuki2013}. Around $\mathbf{q} = \text{(1/2,\,0,\,-1)}$ (M$_1$ point), we observed roton-like minima and a flat mode, such as in the dispersion along $\text{K}'_1$\,$\to$\,M$_1$\,$\to$\,Y$_1$ or X$_1$\,$\to$\,M$_1$\,$\to$\,$\Gamma_1$ [Fig.~\ref{fig:inel}(a)].

The most interesting feature of the magnetic excitations is the line-broadening observed throughout the whole Brillouin zone (BZ) (Fig.~\ref{fig:inel}), which was missed due to the instrument and sample limitations in Ref. \cite{Zhou2012}. As demonstrated in the constant-{\bf q} plots near $\text{M}_1$ point, {\bf q} = (1/2,\, 0\,, -1), the line-widths are several times broader than the instrument resolution [Fig.~\ref{fig:constQ-constE}(a)]. As discussed in Ref.\cite{supplemental}, we have excluded the possible extrinsic broadening factors, such as the sample inhomogeneities, the instrument resolution and data rebinning effects. Similar line-broadening was reported in the 2D trimerized triangular antiferromagnet LuMnO$_{3}$~\cite{Oh2013}. In Fig.~\ref{fig:constQ-constE}(b), we show the constant-frequency cut at $\hbar\omega=1.3\,\text{meV}$ focused on the BZ boundary. Besides the triangular-shaped intensity around K$_1$ and $\text{K}'_1$ points corresponding to a nearly flat single-magnon excitation~\cite{Mourigal2013}, we observed relatively blurred circular-shape intensity. This feature resembles the one observed in the prototypical 2D TLHAF model~\cite{Mourigal2013}. In addition to these broadened quasiparticle peaks, we observed more diffusive features at higher frequency $\hbar\omega \gtrsim 2\,\text{meV}$ [Figs.~\ref{fig:inel}(a)--(c)], which are likely due to the longitudinal spin fluctuations associated with a two-magnon continuum. Finally, we found that the Goldstone mode emanating from the Bragg spot {\bf q} = (2/3,\,-1/3,\,-1) has a rather weak intensity compared to the quadratic gapped branch. Although we believe this is not an experimental artifact, this observation is rather unusual.

\textit{Spin-wave theory.---}%
The INS results were analyzed by the LSW theory and the LSW + $1/S$ corrections  based on the spin-1/2 quasi-2D XXZ Hamiltonian on a vertically stacked triangular lattice:
\begin{align}
  \ham 
  &= J \sum_{\langle{\mathbf{r}^{},\mathbf{r}'}\rangle} 
  \left(
  S_{\mathbf{r}^{}}^{x} S_{\mathbf{r}'}^{x} + S_{\mathbf{r}^{}}^{y} S_{\mathbf{r}'}^{y}
  + \Delta S_\mathbf{r}^{z} S_{\mathbf{r}'}^{z}
  \right)
  \notag\\
  &+ J'\sum_{\mathbf{r}} 
  \left(
  S_{\mathbf{r}^{}}^{x} S_{\mathbf{r}^{}+\hat{z}}^{x} + S_{\mathbf{r}^{}}^{y} S_{\mathbf{r}^{}+\hat{z}}^{y}
  + \Delta S_{\mathbf{r}^{}}^{z} S_{\mathbf{r}^{}+\hat{z}}^{z}
  \right).
  \label{eq:H}
\end{align}
Here $\langle{\mathbf{r},\mathbf{r}'}\rangle$ runs over the in-plane NNs. $J$ and $J'$ ($J,J'>0$) are the intra- and inter-layer NN antiferromagnetic exchange, respectively. $\Delta<1$ parameterizes the easy-plane exchange anisotropy. This is a minimal extension of the isotropic 2D TLHAF model ($J' = 0$ and $\Delta = 1$)~\cite{Zheng2006,Starykh2006,Chernyshev2006,Chernyshev2009,Mourigal2013}. The classical ground state for $J' > 0$ and $\Delta < 1$ coincides with the experimentally observed 120$^\circ$ structure. A tentative rough estimate $J \approx \text{1.6\,meV}$ can be made by comparing the saturation field value (32.8\,T for $\textbf{B} \parallel \hat{\textbf{c}}$~\cite{Susuki2013}) with $g \mu^{}_\text{B} B_\text{sat} = [3 (1 + 2\Delta) J + 2(1 + \Delta)J']S$ assuming $J'/J,\,1 - \Delta \ll 1$ and $g = 3.87$ for $\mathbf{B} \parallel \hat{\textbf{c}}$~\cite{Susuki2013}. This suggests the temperature for our INS measurements to be $T \approx 0.1J$.

The spin-wave theory is derived by the standard Holstein-Primakoff transformation relative to the 120$^\circ$ structure. The LSW dispersion, $\omega_{0}(\textbf{q})$, is obtained by diagonalizing the quadratic part of the spin-wave Hamiltonian. The spiral 120$^\circ$ ordering  leads to three branches of poles in the dynamical spin structure factor at $\omega_{0}(\textbf{q})$ and $\omega_{0}(\textbf{q}\pm\textbf{Q})$ [the solid lines in Figs.~\ref{fig:inel}(d)--\ref{fig:inel}(f)]. Several qualitative features of the spectrum can be already captured at this LSW level. First, the gap ($\approx$\,0.65\,meV) of the quadratic branch at {\bf q} = (2/3,\,-1/3,\,-1) is induced by the easy-plane exchange anisotropy. This gap is proportional to $\sqrt{1 - \Delta}$.  The bandwidth of the Goldstone mode along the [001] direction is $\propto\sqrt{J'/J}$, implying that a rather small value of $J'/J$ can explain the observed bandwidth ($\approx$\,1.1\,meV) [Fig.~\ref{fig:inel}(c)]. These observations remain robust after including the next order corrections in $\text{1}/\textit{S}$. The overlap near  {\bf q} = (2/3,\,-1/3,\,0) of the gapped and gapless  high intensity branches along {\bf q} = (2/3,\,-1/3,\,L) is another characteristic of {\BaCo} [Fig.~\ref{fig:inel}(c)]. 

To quantify the effect of quantum fluctuations on the single-magnon spectrum, we include the next order in $1/S$ to compute the dynamical structure factor $\mathcal{S}(\textbf{q},\omega) = (2\pi N)^{-1}\sum_{\mathbf{r},\mathbf{r}'}\int_{-\infty}^{\infty} dt e^{i[\omega t - \textbf{q}\cdot(\mathbf{r} - \mathbf{r}')]}\langle{\textbf{S}_\mathbf{r}(t) \cdot \textbf{S}_{\mathbf{r}'}(0)}\rangle$ ($N$ is the total number of {\Co} ions) at $T = 0$. The result is shown in Figs.~\ref{fig:inel}(d)--\ref{fig:inel}(f), where the experimental energy resolution ($\approx\text{0.063\,meV}$) has been convoluted \cite{supplemental}. This  quantum correction  arises from the cubic terms that appear in the spin-wave Hamiltonian because of the noncollinear nature of the spin ordering and from a  Hartree-Fock decoupling of the always present quartic terms~\cite{Starykh2006,Chernyshev2006,Chernyshev2009,Zhitomirsky2013,Mourigal2013}. The parameters of our best fitting are $J = \text{1.7\,meV}$, $J'/J = \text{0.05}$, and $\Delta = \text{0.89}$. They are chosen to reproduce the gap of the quadratic branch at $\mathbf{q} = (2/3,-1/3,-1)$, the bandwidth of the Goldstone mode along the [001] direction, and the saturation field for $\textbf{B} \parallel \hat{\textbf{c}}$~\cite{supplemental}. The main difference relative to the previous estimates from ESR measurements ($J'/J = \text{0.026}$ and $\Delta = \text{0.94}$)~\cite{Susuki2013} is the stronger exchange anisotropy. By comparing against the LSW results, we confirm the strong \textit{downward} renormalization [as large as $\approx$\,40\% near {\bf q} = (2/3,\,-1/3,\,-1)] of the single-magnon dispersion. This is a salient feature of frustrated low-spin magnets relative to unfrustrated systems. For instance,  the  renormalization due to $1/S$ corrections is \textit{upward} in the square-lattice $S=1/2$ antiferromagnetic Heisenberg model~\cite{Igarashi2005}. A rather small  downward renormalization of $\approx$\,5\% is observed  in the $S=2$ triangular lattice compound LuMnO$_3$ because of the rather large value of the spin~\cite{Oh2013}.

While many key features of the measured spectra are well captured by our minimal model, there are several noteworthy discrepancies. First, as expected,  the calculation yields high intensity for the Goldstone mode. Second, while the shape of the low-frequency dispersion  ($\lesssim$\,1.5\,meV) is well reproduced, the calculation overestimates the energy of the high-frequency part (e.g., $\approx$\,2.3\,meV near $\text{M}_1$ point whereas it is $\approx$\,1.7\,meV experimentally). This overestimate is a robust feature of our minimal Hamiltonian~\eqref{eq:H} at both LSW and LSW + $1/S$ levels~\cite{supplemental}. We examined the effect of the antiferromagnetic next-nearest-neighbor intralayer exchange $J_2$, which however lowers both energy scales and relatively speaking, the separation between the two branches increases with $J_2$~\cite{supplemental}.

More importantly, the LSW + $1/S$ calculation yields quite stable quasiparticle peaks [Figs.~\ref{fig:inel}(d)--\ref{fig:inel}(f)] instead of the observed line-broadening in the parameter regime relevant for {\BaCo} [Figs.~\ref{fig:inel}(a)--\ref{fig:inel}(c)]. The large $S$ theory applied to the 2D isotropic TLHAF model predicts that single magnon excitations can decay into two-magnon continuum in a large region of the BZ where the kinematic conditions are satisfied~\cite{Starykh2006,Chernyshev2006,Chernyshev2009,Zhitomirsky2013,Mourigal2013}. Such magnon decays make the magnon lifetime finite and cause the line-broadening. However, the nonnegligible easy-plane exchange-anisotropy in {\BaCo} implies that this semiclassical scenario fails to reproduce the experimental observation because, as already pointed out in Refs.~\cite{Chernyshev2006,Chernyshev2009}, the associated gap opening violates the kinematic condition in the increasingly large area of the BZ~\cite{supplemental}. In addition, the inter-layer coupling also acts against spontaneous magnon decays. One possibility is that we are missing some significant interactions in our minimal Hamiltonian~\eqref{eq:H}. An alternative explanation is that semiclassical approaches are simply inadequate to describe magnon decay in low-dimensional frustrated spin systems with small $S$.

\textit{Conclusions.---}%
In summary, our neutron diffraction measurements of single-crystal {\BaCo}  confirm that the zero-field magnetic ordering is a 120$^\circ$ structure in the $ab$ plane. By comparing our measurements against the dynamical spin structure factor obtained from the LSW + $1/S$ treatment of a stacked triangular-lattice $S=1/2$ XXZ model, we extracted both exchange and anisotropy parameters. Our results indicate that  {\BaCo} is an almost ideal realization of an equilateral $S=1/2$ TLHAF. The measured INS profile exhibits several salient features theoretically predicted for frustrated quantum magnets, such as the strong downward renormalization of the magnon dispersion, roton-like minima, flat modes near the BZ boundary, and the line-broadening throughout the entire BZ. However, our detailed comparison between the experiments and the large $S$ treatments reveals that the observed magnon decay in {\BaCo} cannot be explained with a semi-classical treatment. Thus our study suggests that  a new theoretical framework is need to describe the low-energy excitation spectrum of magnetically ordered low-dimensional frustrated magnets.

\begin{acknowledgments}
  \textit{Acknowledgments.---}%
  The authors acknowledge valuable discussions with M. Mourigal. The research at HFIR and SNS at ORNL were sponsored by the Scientific User Facilities Division (J.M., T.H., H.B.C., W.T., and M.M.), Office of Basic Energy Sciences, US Department of Energy. J.M., Z.L.D.~and H.D.Z.~acknowledge the support from NSF-DMR through Award DMR-1350002 and from NHMFL through No. NSF-DMR-1157490, U.S. DOE, and the State of Florida. Y.K.~acknowledges financial supports from the RIKEN iTHES project.  Work at LANL was performed under the auspices of the U.S.\ DOE contract No.~DE-AC52-06NA25396 through the LDRD program.
\end{acknowledgments}

\section{NONLINEAR SPIN-WAVE ANALYSIS}

We summarize the results of the nonlinear spin-wave (LSW + $1/S$) calculation~\cite{Starykh2006,Chernyshev2006,Chernyshev2009,Mourigal2013} of the dynamical spin structure factor for the quasi-2D XXZ model on the stacked triangular lattice.
In addition to the easy-plane exchange anisotropy and the inter-layer coupling, we examine the effect of the intra-layer next-nearest-neighbor (NNN) coupling $J_2$. The extended Hamiltonian is
\begin{align}
  \ham_\text{\,$J$-$J_2$-$J'$}^{}
  &= J \sum_{\langle{\mathbf{r}^{},\mathbf{r}'}\rangle} 
  \left(
  S_{\mathbf{r}^{}}^{x} S_{\mathbf{r}'}^{x} + S_{\mathbf{r}^{}}^{y} S_{\mathbf{r}'}^{y}
  + \Delta S_\mathbf{r}^{z} S_{\mathbf{r}'}^{z}
  \right)
  \notag\\
  &+ J_2 \sum_{\langle{\mathbf{r}^{},\mathbf{r}'}\rangle_2} 
  \left(
  S_{\mathbf{r}^{}}^{x} S_{\mathbf{r}'}^{x} + S_{\mathbf{r}^{}}^{y} S_{\mathbf{r}'}^{y}
  + \Delta S_\mathbf{r}^{z} S_{\mathbf{r}'}^{z}
  \right)
  \notag\\
  &+ J'\sum_{\mathbf{r}} 
  \left(
  S_{\mathbf{r}^{}}^{x} S_{\mathbf{r}^{}+\hat{z}}^{x} + S_{\mathbf{r}^{}}^{y} S_{\mathbf{r}^{}+\hat{z}}^{y}
  + \Delta S_{\mathbf{r}^{}}^{z} S_{\mathbf{r}^{}+\hat{z}}^{z}
  \right),
  \label{sppl:eq:H}
\end{align}
where $\langle{\mathbf{r}^{},\mathbf{r}'}\rangle$ and $\langle{\mathbf{r}^{},\mathbf{r}'}\rangle_2$ run over nearest neighbor and NNN pairs in each layer, respectively.

\begin{figure}
  \centering
  \includegraphics[width=0.8\hsize]{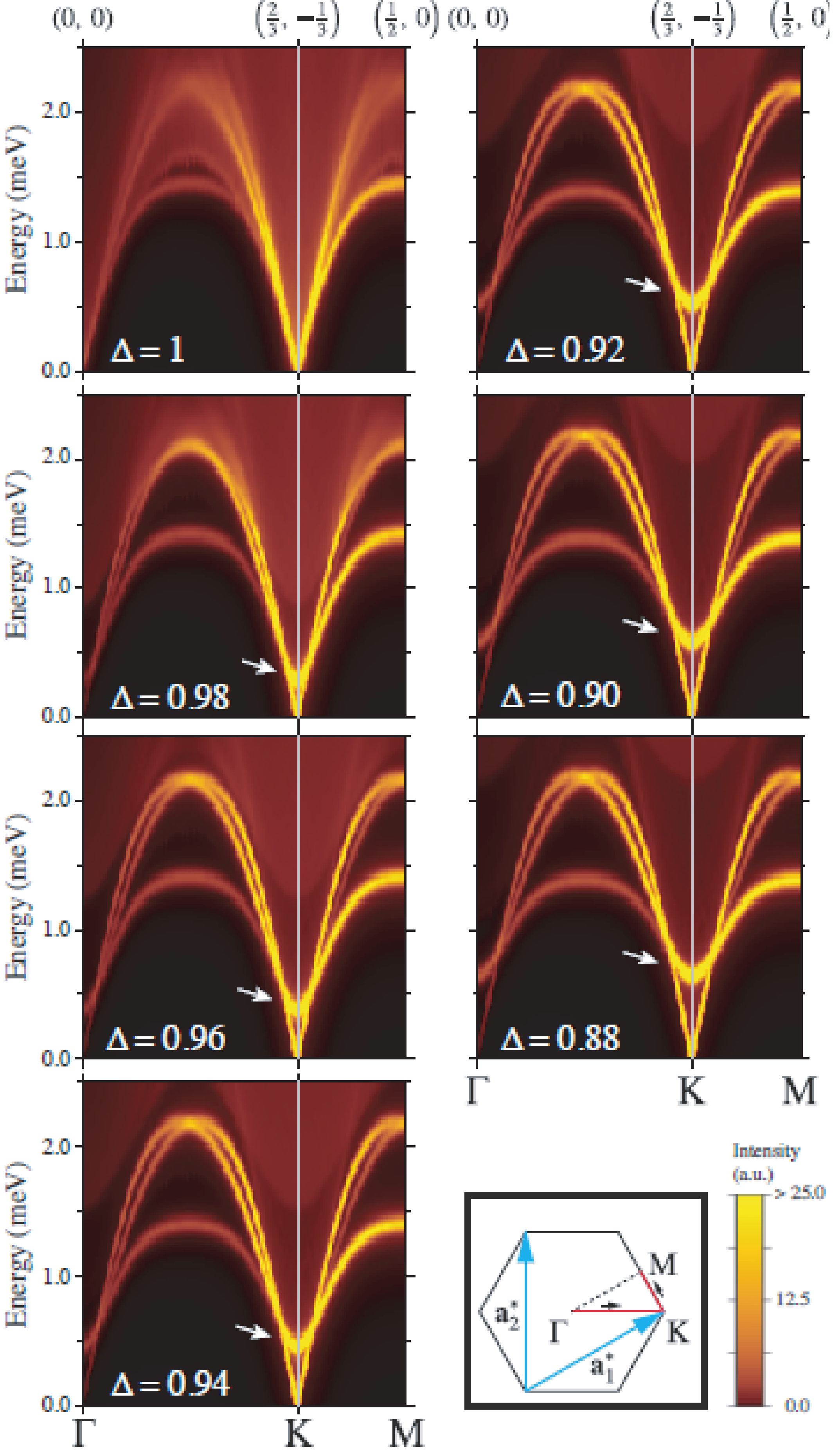}
  \caption{\label{fig:NN-2D}%
    Dynamical spin structure factor in the 2D nearest-neighbor XXZ model at $T = 0$ for selected values of the easy-plane anisotropy $\Delta$. The energy resolution $\hbar \delta \omega \approx 0.063\,\text{meV}$ in our neutron scattering measurement has been convoluted. The arrows indicate the quadratic branch at K point for $\Delta < 1$. $J = 1.7\,\text{meV}$ is assumed. }
\end{figure}

\begin{figure*}
  \centering
  \includegraphics[width=\hsize]{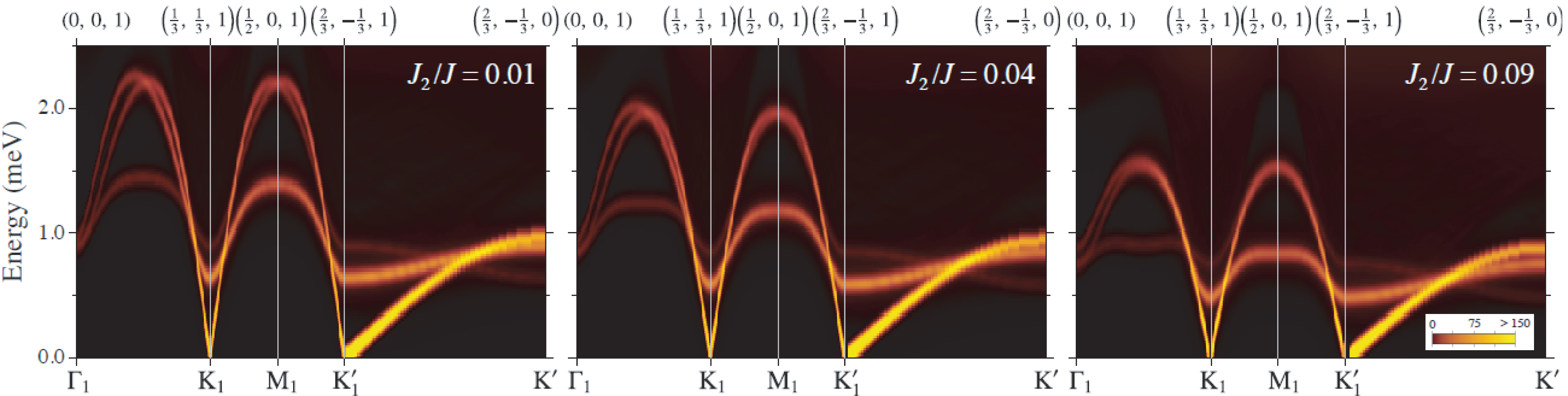}
  \caption{\label{fig:NN-Q2D}%
    Comparisons between $J'/J = 0.01$ (top row) and $J'/J = 0.04$ (bottom row) for selected values of the exchange anisotropy: $\Delta = 0.96$, $0.92$, and $0.88$. The inset shows the first Brillouin zone. The energy resolution of $\hbar \delta \omega \approx 0.063\,\text{meV}$ has been convoluted.
  }
\end{figure*}

\subsection{Nearest-neighbor $XXZ$ model in the 2D limit}
We focus on the parameter dependence of (i) the line broadening of the single-particle dispersion and (ii) its renormalization, especially near the top of the dispersions.
We begin with the case of $J' = J_2 = 0$ and examine the effect of $\Delta < 1$ at $T = 0$. In Fig.~\ref{fig:NN-2D}, we show the $\Delta$-dependence of the dynamical spin structure factor. Our results demonstrate that the line broadening is significantly suppressed throughout the whole Brillouin zone for $\Delta \approx 0.89$, which is required to explain the sizable gap $\approx\,\text{0.65\,meV}$ (around 40\% of $J$) of the quadratic band at the K$_1$ and K$'_1$ points in {\BaCo}.
As discussed in Refs.~\cite{Chernyshev2006,Chernyshev2009}, the line broadening evident in the isotropic case $\Delta = 1$ is largely due to the magnon decay process involving the gapless $\mathbf{k} = \pm\mathbf{Q}$ magnons, which however become gapped for $\Delta < 1$. This implies that, as $\Delta$ decreases, decays with emission of $\mathbf{k} = \pm\mathbf{Q}$ magnons are kinematically prohibited in an increasingly large region of the Brillouin zone~\cite{Chernyshev2006,Chernyshev2009}. Thus, the sizable easy-plane exchange anisotropy required to explain the observed gap is incompatible with the observed line broadening within the spin-wave approach.

\begin{figure*}
 \centering
 \includegraphics[width=1.0\textwidth]{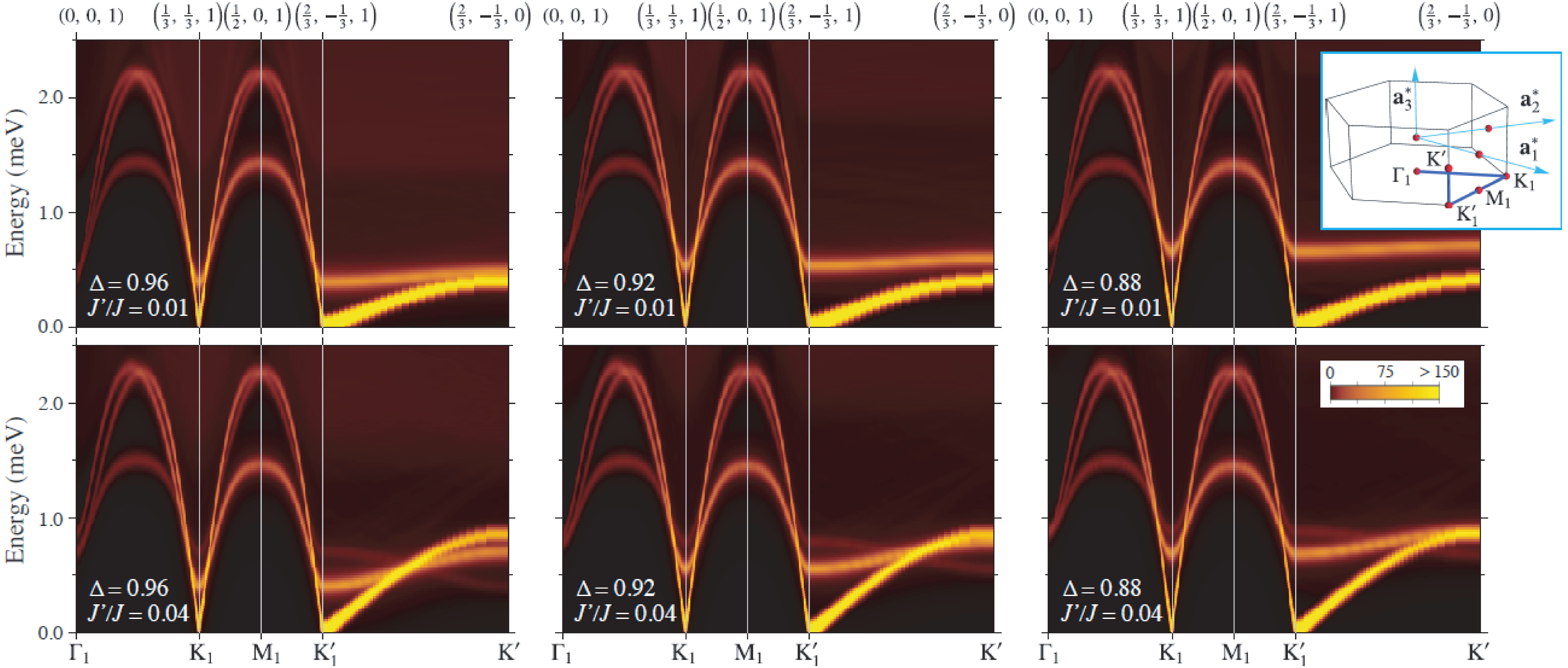}
  \caption{The $J_2$-dependence of the dynamical spin structure factor at $T = 0$ ($\Delta = 0.89$ and $J'/J = 0.05$).  The energy resolution of $\hbar \delta \omega \approx 0.063\,\text{meV}$ has been convoluted.}
  \label{diff_J2}
\end{figure*}

\subsection{Quasi-2D nearest-neighbor $XXZ$ model}
Next, we examine the effect of the antiferromagnetic inter-layer exchange $J' > 0$ (Fig.~\ref{fig:NN-Q2D}). As expected, the dominant effect of $J' > 0$ appears in the dispersion along the inter-layer direction. For small $J'/J$, the bandwidth of the Goldstone mode and the quadratic mode along the third direction scale as $\propto \sqrt{J'/J}$ and $J'$, respectively. As a consequence, at a certain value of $J'/J$, these modes touch to each other at the K and $\text{K}'$ points [$\mathbf{q} = (1/3,1/3,0),\,(2/3,-1/3,0)$], as observed in {\BaCo}. As can be seen in Fig.~\ref{fig:NN-Q2D}, the intrinsic line broadening is basically absent in the whole Brillouin zone.

Meanwhile, we note that a Schwinger bosons mean field approach yields a spectrum with broadening due to deconfined spinons~\cite{Ghioldi2015}, bearing some resemblance to the observed results. After extending this approach to discuss spinon confinement as a more reliable description of a magnetically ordered state, it may offer a way of reconciling the discrepancy between theory and experiments.

\subsection{Effect of $J_2$}
The other major discrepancy between our theory and experimental observations in {\BaCo} is about the distance between tops of different magnon branches [see, e.g., Figs.~2(b) and 2(e) in the main text]. Theoretically, the LSW approximation predicts the frequencies at the M$_1$ point (corresponding to the M point in 2D) to be $\sqrt{5/2}J$ and $J$ for the 2D nearest-neighbor isotropic TLHAF. On the other hand, our INS measurements show that these frequencies are $\approx$\,1.7\,meV and $\approx$\,1.3\,meV in {\BaCo}, and the corresponding ratio $\approx$\,1.3 between them is considerably smaller than the theoretical value. As can be seen in Figs.~\ref{fig:NN-2D} and \ref{fig:NN-Q2D}, the change of the ratio between the two due to $\Delta < 1$, $J' > 0$, and the lowest-order $1/S$ corrections is rather small. 
In Fig.~\ref{diff_J2}, we also examine the effect of the antiferromagnetic intra-layer NNN exchange, $J_2$, to see whether this discrepancy can be reduced by introducing an additional source of frustration. However, $J_2$ renormalizes both branches downwards resulting in a ratio that is even higher than the one obtained for $J_2 = 0$.

\section{Determination of Site-disorder by THE SINGLE CRYSTAL NEUTRON DIFFRACTION }

\begin{figure}
 \centering
 \includegraphics[width=0.4\textwidth]{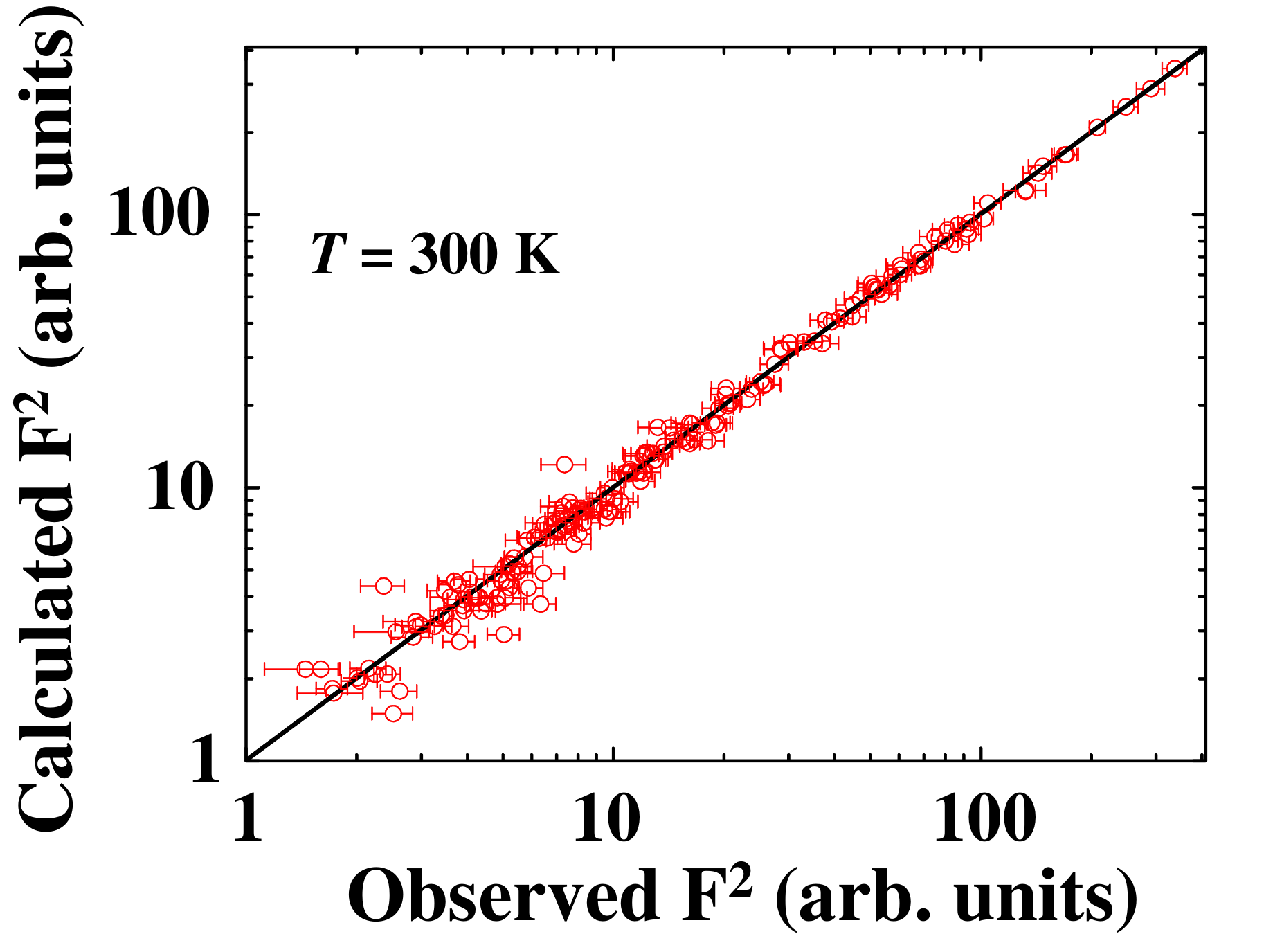}
  \caption{Comparison between the observed  intensities, corresponding to the squared nuclear structure factors, at 300$\,$K on the HB-3A and the calculated ones based on the hexagonal structure with a symmetry of, $P$6$_3/mmc$. The solid line is a guide to the eye.}
  \label{diff_comp}
\end{figure}

To exclude the possibility that the observed linewidth broadening is caused by sample inhomogeneity, we collected more than 200 inequivalent nuclear Bragg peaks on the Four-Circle Diffractometer, HB3A at HFIR, ORNL with the neutron wavelength of $\lambda=1.003 \AA$ at room temperature. The Rietveld refinement was carried out using the FULLPROF program, Fig.\ref{diff_comp}. The $site$-disorder between Co$^{2+}$ and Sb$^{5+}$ ions is determined to be negligible with the standard deviation of 1$\%$. 

\section{Instrumental resolution and data binning effect}

Another possibility to cause the linewidth broadening is the instrumental resolution and data binning effect.

Since the dispersion modes are flat at the band top, the q-resolution effect is small and the energy-resolution should be dominant at the high energy region(above $\sim\text{1.0\,meV}$). Although we have convoluted the calculated dispersions with an energy resolution of $\text{0.063\,meV}$, which is the energy resolution at the elastic line, the measured dispersions are still far broader than the calculated ones. The instrumental resolution does not cause the significant broadening.

In order to check the data binning and integration effects, we have performed data analysis with several different binning/integration conditions: The two excitations at M$_1$-point do not change the linewidth with the different smoothing levels. Hence, the binning and integration effects do not cause the broadening. 

Comparing to the other classical triangular-lattice antiferromagnets with noncollinear magnetic structure, such as CuCrO$_2$ ~\cite{Frontzek2011}, and CuFe$_{0.965}$Ga$_{0.035}$O$_2$~\cite{Nakajima2012}, the linewidths of Ba$_3$CoSb$_2$O$_9$ are broader, which is due to the small spin value ($S$ = 1/2) and the related strong quantum effect.

\bibliographystyle{apsrev}

\end{document}